\documentclass[11pt]{iopjournal}
\usepackage{graphicx,amssymb,amsmath,color}
\usepackage{xcolor}
\usepackage{hyperref}
\hypersetup{
    colorlinks=true,
    linkcolor=blue,
    citecolor=blue,
    urlcolor=blue
}
\usepackage{cite}
\usepackage{nccmath}
\usepackage{comment}
\usepackage[normalem]{ulem} 
\usepackage{color}
\begin{document}
\title{Diffraction induced quantum chaos \\in a one-dimensional Bose gas}
\author{M. Olshanii$^1$, G. Aupetit-Diallo$^2$, S.~G. Jackson$^3$, P. Vignolo$^{4,5}$, and M. Albert$^{4,5,*}$}

\affil{$^1$Department of Physics, University of Massachusetts Boston, Boston, Massachusetts 02125, USA}

\affil{$^2$SISSA, Via Bonomea 265, I-34136 Trieste, Italy}

\affil{$^3$Department of Mathematics, University of Massachusetts Boston, Boston, Massachusetts 02125, USA }

\affil{$^4$Universit\'e C\^ote d'Azur, CNRS, Institut de Physique de Nice, 06200 Nice, France}

\affil{$^5$Institut Universitaire de France}

\affil{$^*$Corresponding author: \href{mailto:mathias.albert@univ-cotedazur.fr}{mathias.albert@univ-cotedazur.fr}}

\begin{abstract}
  We investigate the Lieb--Liniger model of interacting one-dimensional bosons coupled to a localized impurity, modeled by a delta barrier. While the Lieb--Liniger gas is integrable, the impurity breaks integrability and induces a transition towards quantum chaos. We show that the low-energy spectrum exhibits random-matrix statistics, in striking contrast to the Bohigas--Giannoni--Schmit conjecture, where chaotic behavior typically emerges at high energy. For two bosons, the odd-parity sector remains integrable, whereas the even-parity sector displays clear signatures of chaos at low energy and a crossover back to quasi-integrable behavior at higher energies. For three bosons, both parity sectors exhibit spectral statistics close to chaos at low energy. We argue that this unconventional form of few-body quantum chaos originates from diffractive processes induced by the impurity.
\end{abstract}

\section{Introduction}
\medskip
The study of quantum chaos seeks to understand how signatures of classical chaotic dynamics manifest in quantum systems. Since the pioneering works on classically chaotic systems, spectral statistics have emerged as one of the most powerful tools to diagnose quantum chaos. In particular, the nearest-neighbor level spacing distribution (LSD) distinguishes between integrable and chaotic dynamics: integrable systems typically display uncorrelated spectra characterized by a Poisson distribution, while chaotic systems exhibit strong level repulsion described by random matrix theory (RMT). Indeed, according to the celebrated Bohigas--Giannoni--Schmit (BGS) conjecture \cite{BGS1984}, the spectral statistics of a generic quantum system whose classical counterpart is chaotic follow those of Gaussian ensembles, whereas integrable systems show Poissonian behavior. Although exceptions are known, this paradigm has provided the main framework to classify spectral properties of complex quantum systems ranging from nuclei and atoms to mesoscopic structures \cite{RMTreview,RMTreview2}.

A particularly rich arena for exploring quantum chaos is offered by many-body one-dimensional (1D) integrable systems supplemented by a weak integrability-breaking perturbation. Among the most celebrated integrable 1D models is the Lieb--Liniger (LL) model \cite{LiebLiniger}, describing bosons with contact interactions. The LL model is exactly solvable by the Bethe Ansatz and possesses an infinite number of conserved quantities, which render it integrable. As a consequence, it does not thermalize, but instead exhibits constrained dynamics captured by generalized Gibbs ensembles. From an experimental perspective, the LL model provides an excellent description of ultracold atomic gases in tight waveguides \cite{Paredesetal,Kinoshita2004,Kinoshita2005,CazalillaCitroGiamarchiOrigancRigol}, where interactions can be tuned with high precision. From a theoretical point of view, its exact solvability makes it a paradigmatic model for the study of non-equilibrium dynamics and integrability.

The cornerstone of Bethe Ansatz solvability is the absence of diffraction in scattering processes. With the continuum of scattered momenta being suppressed, the set of the particle momenta that remains prominent is identical to the one in the classical counterpart, the latter being finite. The latter fact promotes integrability. Conversely, when diffraction is present, integrability is lost and chaotic behavior may emerge. A natural way to break diffractionless scattering---and hence integrability---of the LL model is to introduce a localized impurity potential, such as a delta-function barrier \cite{WangProsen2025}.

In this work, we investigate what happens to the spectral statistics of the LL model when a barrier is inserted. Using exact diagonalization for few-body systems, we show that diffraction induced by the impurity leads to the emergence of chaos at low energy. For two bosons, the odd-parity sector remains integrable, while the even-parity sector exhibits random-matrix statistics at low energy and a crossover towards quasi-integrable behavior at higher energies. For three bosons, both parity sectors display signatures of chaos at low energy. This unconventional scenario, where chaos emerges in the lowest part of the spectrum rather than at high energy as suggested by the BGS conjecture, highlights the central role of diffraction in shaping the spectral properties of interacting quantum systems. Our results shed new light on the mechanisms by which integrability can be destroyed in a controllable way in ultracold gases, with direct implications for their dynamical and thermalization properties.

The paper is organized as follows. In section \ref{sec:model} we introduce the model under investigation and the method to solve it. In sections \ref{sec:results_2p} and \ref{sec:results_3p} we present our results for two and three particles respectively. Section \ref{sec:discussion} is devoted to a general discussion about diffraction induced quantum chaos and then we conclude in Sec. \ref{sec:conclusion}. Technical details about the calculation of the matrix elements are reported in App. \ref{sec:appendix}.
\section{Model \label{sec:model}}
\medskip
We consider $N$ identical bosons of mass $m$ and coordinates $x_i$, living on a ring of size $L$ with point-like repulsive interactions of strength $g$, and periodic boundary conditions. In addition, the atoms are subjected to a local potential of the form $g_B\delta (x_i)$.
\begin{equation}\label{eq_H_LL}
  H = -\frac{\hbar^2}{2m}\sum_{i=1}^N \frac{\partial^2}{\partial x_i^2} + g \, \sum_{i>j} \delta(x_i-x_j)+g_B \sum_{i=1}^N \delta(x_i).
\end{equation}

When $g_B$ is set to zero, we recover the standard LL Hamiltonian which can be exactly solved with the Bethe Ansatz \cite{LiebLiniger}. The eigenstates are labeled by a set of $N$ rapidities $\lambda_i$ and can be written, in the fundamental sector $x_1\le x_2\le ...\le x_N$, as
\begin{equation}\label{eq_psi}
  \Psi_{\{\lambda_j\}}(\{x_j\}) = \sum_{P\in S_N} A_P \exp\Bigl(i \sum_{k=1}^N \lambda_{P(k)} x_k\Bigr),
\end{equation}
where $S_N$ is the permutation group of $N$ elements. The $A_P$ are coefficients given by
\begin{equation}\label{eq_ap}
  A_P = \mathcal N (-1)^P \prod_{1\le k < j \le N}[\lambda_{P(j)} - \lambda_{P(k)}-ic]
\end{equation}
with $(-1)^P$ the signature of permutation $P$, $\mathcal N$ the normalization constant and $c=gm/\hbar^2$. In addition, the rapidities are all real and fulfill the $N$ coupled Bethe equations
\begin{equation}\label{eq_bethe}
  \lambda_j = \frac{2\pi}{L} I_j - \frac{2}{L} \sum_{k=1}^N \textrm{arctan}\left(\frac{\lambda_j - \lambda_k}{c}\right),
\end{equation}
where the Bethe numbers $I_j$ are integers for odd $N$ and half-integers for even $N$, and must all be distinct. The energy of a Bethe state $|\vec \lambda\rangle$ is given by $E_{\vec \lambda} = \frac{\hbar^2}{2m} \sum_{i=1}^N \lambda_i^2$ and its momentum $P_{\vec \lambda} = \hbar \sum_{i=1}^N \lambda_i$. The ground state corresponds to the lowest available set of centered Bethe numbers. For instance, for $N=2$ it is given by $\vec I=(-1/2,+1/2)$ and for $N=3$ by $\vec I=(-1,0,+1)$.
Due to the bosonic statistics, these states are symmetric under permutations of spatial coordinates and rapidities, allowing us to restrict our calculations to the fundamental rapidity sector $I_1<I_2<...<I_N$. It is important to note that the rapidities $\lambda_i$ differ from the actual momenta of the particles except in the non-interacting limit. In the limit of infinite interactions ($c \to +\infty$), the rapidities coincide with the momenta of the corresponding free fermions via the Bose-Fermi mapping \cite{Girardeau}. In general, however, they do not reside on a regular lattice (which would have spacing $ 2\pi/L$), but are instead coupled through the Bethe equations \eqref{eq_bethe}.

In order to diagonalize the Hamiltonian (\ref{eq_H_LL}), we will use the Bethe basis introduced above. The matrix elements of the impurity potential can be computed analytically and the resulting Hamiltonian is then diagonalized numerically within a truncated Hilbert space. These matrix elements are known in the literature on integrable systems as form factors \cite{Korepinetal,Slavnov,deNardisPanfil}, since they reduce to the matrix elements of the density operator at $\vec{x}=\vec 0$. Details of the numerical procedure are provided in the following sections and in the appendix \ref{sec:appendix}.

This approach gives us direct access to both the spectrum and the eigenstates, from which we extract the LSD as well as properties of the eigenstates such as the participation ratio. The latter gives information about the number of basis states that participate in a given eigenstate. The eigenstates are expanded in the Bethe basis as $|\psi_n\rangle=\sum_{\vec \lambda} \alpha^{(n)}_{\vec\lambda} |\vec\lambda\rangle$.

Within these notations, the participation ratio is defined as
\begin{equation}\label{eq_PR}
  P_n=\left(\sum_{\vec\lambda} |\alpha^{(n)}_{\vec \lambda}|^4\right)^{-1}.
\end{equation}

The spectral statistics are obtained from unfolded numerical spectra in different parity sectors. These parity sectors are defined by $\langle -\vec x | \psi_n\rangle=\pm \langle \vec x | \psi_n\rangle$. We compare our results with the well known limiting cases referred to as Poisson, $p(s)=\exp(-s)$ or Wigner--Dyson for the Gaussian Orthogonal Ensemble $p(s)=\frac{\pi}{2} s \exp[-\frac{\pi}{4} s^2]$ as well as the Brody distribution \cite{brody1973} $(\beta+1)bs^\beta \exp[-bs^{\beta+1}]$ with $b=[\Gamma(\beta+2)/(\beta+1)]^{\beta+1}$ where $\Gamma$ is the Euler Gamma function. The Brody distribution reduces to the Poisson or Wigner--Dyson ones for $\beta = 0$ or $1$ respectively.

Before discussing our results, it is important to recall the spectral properties in the integrable limit $g_B = 0$. This case has been extensively analyzed in Ref.~\cite{Izrailev2021} in the context of the long-standing question of spectral statistics in integrable quantum systems. In this limit, the statistical properties of the many-body spectrum are highly nontrivial and depend crucially on whether one considers the full spectrum or a restricted subset with fixed total momentum. Since total momentum is conserved in the absence of impurities, mixing all momentum sectors leads to strong level clustering, making it meaningless to speak of a Poissonian LSD. Such a distribution emerges only within a fixed momentum sector, and only for intermediate interaction strengths. In our study of the LL gas coupled to a localized impurity, total momentum is no longer conserved and parity becomes the relevant quantum number. We therefore focus on the parameter regime where, in the absence of the impurity, the spectrum restricted to fixed momentum exhibits Poissonian statistics, providing a clean reference point to identify the onset of chaos induced by the barrier. Figure~\ref{fig_spectrum_2p} (left) provides a schematic overview of our main results. The spectrum separates into a low-energy chaotic region with Wigner--Dyson statistics and a high-energy quasi-integrable region displaying Poissonian statistics. The crossover energy between these two regimes depends nontrivially on the system parameters but is mainly set by the barrier strength.

\section{Results for two particles\label{sec:results_2p}}
\medskip
We now present our results for the case of $N=2$ bosons as functions of the rescaled interaction strengths $\gamma=gmL/(\hbar^2 N)=cL/N$ and
$\gamma_B=g_BmL/(\hbar^2 N)$. (In the numerical calculations, we used a system of units $\hbar^2/m =1 $ and $L=N$.)

\begin{figure}
  \centering\includegraphics[width=0.7\linewidth]{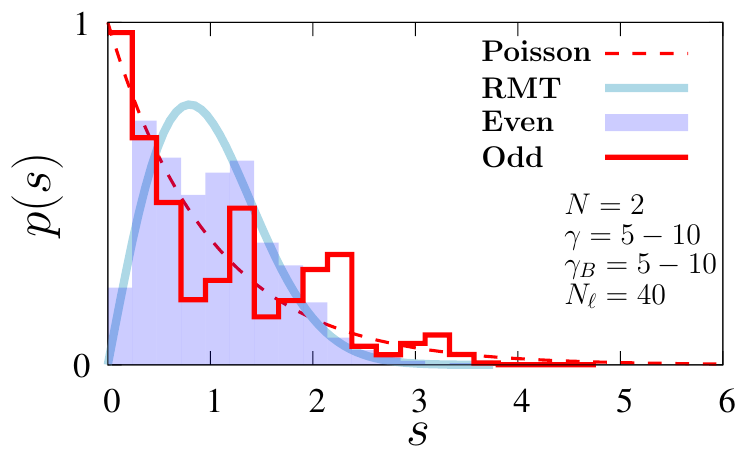}
  \caption{Level spacing distribution of the unfolded spectrum for two particles. The histogram is computed over the lowest $N_\ell=40$ energy levels for different $\gamma$ and $\gamma_B$ in the range [5-10]. In total 4400 energy levels are used to compute these histograms. The levels are sorted by parity and the numerical histograms are compared to the Poisson and Wigner--Dyson distributions.\label{fig_ps_2p} }
\end{figure}

\begin{figure}
  \centering\includegraphics[width=0.7\linewidth]{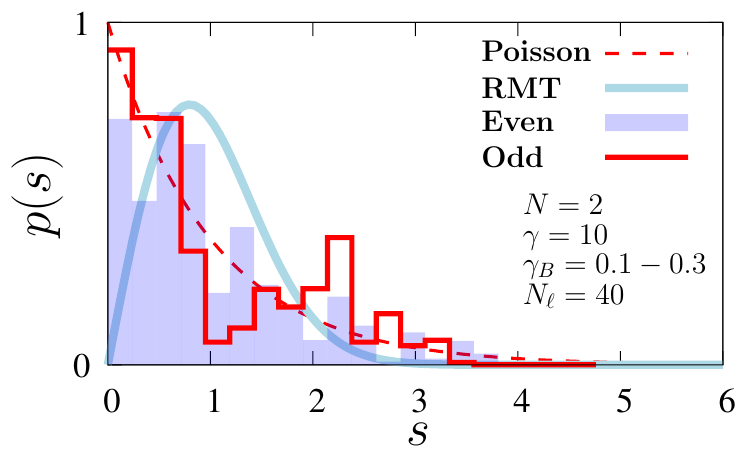}
  \caption{Level spacing distribution of the unfolded spectrum for two particles. The histogram is computed over the lowest $N_\ell=40$ energy levels for $\gamma=10$ and $\gamma_B$ in the range [0.1-0.3]. In total 1000 energy levels are used to compute these histograms. The levels are sorted by parity and the numerical histograms are compared to the Poisson and Wigner--Dyson distributions.\label{fig_ps_2p_2} }
\end{figure}

\begin{figure}
  \centering\includegraphics[width=0.7\linewidth]{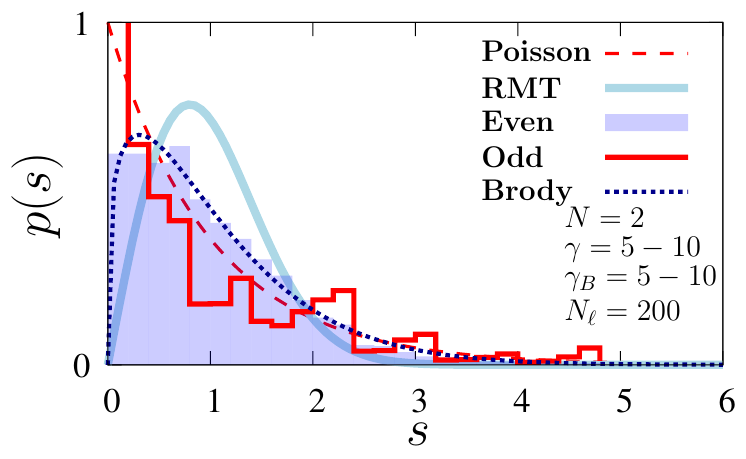}
  \caption{Level spacing distribution of the unfolded spectrum for two particles. The histogram is computed over the lowest $N_\ell=200$ energy levels for different $\gamma$ and $\gamma_B$ in the range [5-10]. In total 22000 levels are used to compute these histograms. The levels are sorted by parity and the numerical histograms are compared to the Poisson, Wigner--Dyson and Brody distributions (with $\beta=0.2505$) \cite{fitbrody}.\label{fig_ps_2p3} }
\end{figure}

It is known that the odd-parity sector of the spectrum can be obtained through the asymmetric Bethe Ansatz method \cite{Jackson2025} and remains integrable. The exact diagonalization, however, allows us to compute the entire spectrum. We use a basis constructed from Bethe states with quantum numbers $-N_s/2 \le I_1 < I_2 \le N_s/2$ (half-integers), with $N_s = 101$. The resulting truncated Hilbert space has dimension $M = \binom{N_s}{2} = 5050$. We have verified that our results remain stable upon increasing the number of basis states, and that in the odd-parity sector we recover the spectrum predicted in Ref.~\cite{Jackson2025}.
\begin{figure}
  \centering
  \includegraphics[width=0.35\linewidth]{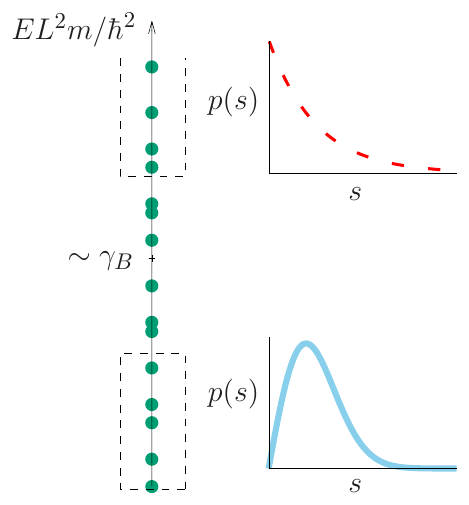}
  \includegraphics[width=0.6\linewidth]{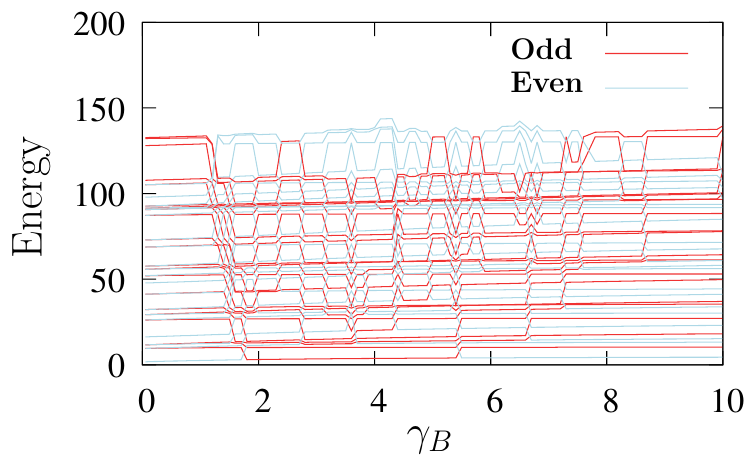}
  \caption{Left: Scheme of the energy spectrum and the corresponding level spacing distribution. The frontier between the low and high energy parts of the spectrum is located at an energy of the order of $\gamma_B$. Right: Actual energy spectrum (first 20 levels in units of $\hbar^2/mL^2$) for two particles and $\gamma=10$ as a function of the barrier coupling strength $\gamma_B$.\label{fig_spectrum_2p}
  }
\end{figure}
In Fig.~\ref{fig_spectrum_2p} we show the first twenty energy levels as a function of the effective impurity coupling $\gamma_B$ for a fixed effective interaction strength $\gamma = 10$. In the range $\gamma_B \in [2,10]$, the spectrum develops a complex structure characterized by level crossings between states of different parity and avoided crossings between states of even parity. It is in this regime that we expect signatures of quantum chaos, as revealed by the LSD. The latter is displayed in Fig.~\ref{fig_ps_2p}, where it is computed for different values of the interaction and impurity parameters within this range in order to improve statistics. The two parity sectors are analyzed separately and display, respectively, Poissonian statistics characteristic of integrable systems in the odd sector, and Wigner--Dyson statistics characteristic of random matrix theory in the even sector. The behavior near zero spacing provides a particularly sharp distinction between integrability and quantum chaos in both cases. On the other hand, Fig.~\ref{fig_ps_2p_2} shows that away from this regime, the statistics becomes Poissonian in both cases. Finally, when higher energy states are considered, the LSD of even states deviates from Wigner--Dyson and follows a Brody distribution closer to Poisson as shown on Fig~\ref{fig_ps_2p3}.

\begin{figure}
  \centering\includegraphics[width=0.6\linewidth]{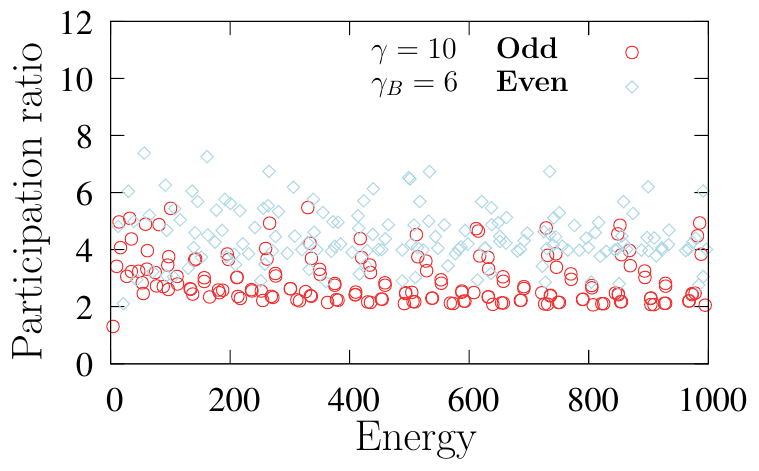}
  \caption{Participation ratio of the eigenstates of hamiltonian (\ref{eq_H_LL}) for $N=2$, $\gamma=10$ and $\gamma_B=6$. Red circles correspond to the odd parity states and blue diamonds to the even parity states. Energy is in units of $\hbar^2/mL^2$.\label{fig_PR_2p}}
\end{figure}

To gain further insight into the difference between low- and high-energy states, we computed the participation ratio, shown in Fig.~\ref{fig_PR_2p}. Even-parity states generally display a much larger participation ratio than odd-parity states. However, the participation ratio does not increase systematically with energy (or state index), as would be expected for quantum-ergodic eigenstates. Ergodic states are indeed expected to spread uniformly over the constant-energy shell, so that the participation ratio---roughly measuring the number of basis states needed to construct an eigenstate---should grow proportionally with the size of this shell. Instead, the participation ratio appears to saturate, which explains why high-energy states do not exhibit chaotic behavior. This can be understood from the fact that the matrix elements of the barrier are bounded by $\gamma_B$ times the density at the impurity position. At high energies, the coupling to the barrier thus becomes negligible, since the kinetic energy of the particles largely exceeds the energy scale set by the impurity.

\begin{figure}
  \includegraphics[width=0.49\linewidth]{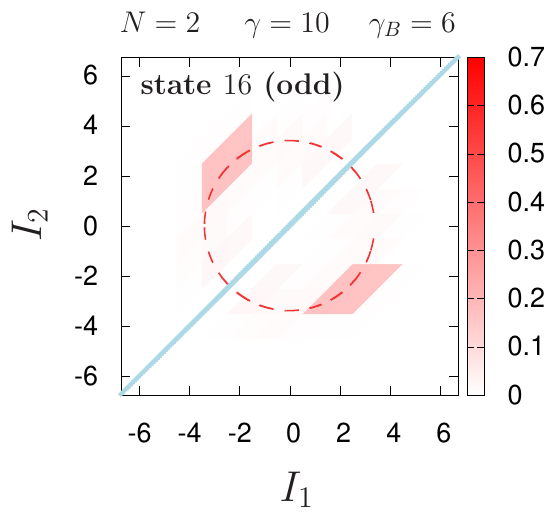}
  \includegraphics[width=0.49\linewidth]{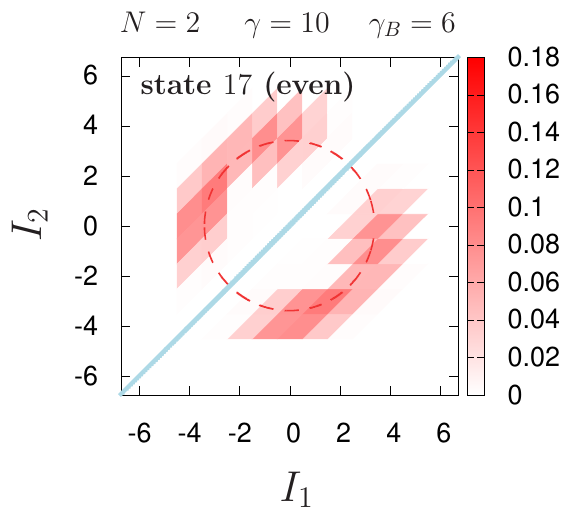}
  \caption{Modulus square of the wave function $|\alpha^{(n)}_{\vec \lambda}|^2$ for two neighboring states in energy but with different parity. State $n=16$ is odd and is mainly localized on a single Bethe state. State $n=17$ is even and spreads uniformly over the constant energy shell defined as $E=\frac{\hbar^2}{2m}\left(\frac{2\pi}{L}\right)^2 (I_1^2+I_2^2)$ (red dashed line) in Bethe space (here $E\simeq 55\hbar^2/mL^2$). The gray line displays the bosonic symmetry line under permutation of Bethe numbers. \label{fig_states_2p}}
\end{figure}

Finally, in Fig.~\ref{fig_states_2p} we show two representative eigenstates in the low-energy sector. These states are nearly degenerate in energy but differ in parity. The modulus of the wavefunction coefficients, $|\alpha^{(n)}_{\vec{\lambda}}|^2$, is plotted in the Bethe-number plane $(I_1,I_2)$. Remarkably, the odd-parity state is strongly localized on a single Bethe basis state, while the even-parity state is spread almost uniformly over the constant-energy shell---a hallmark of a quantum-ergodic eigenstate. Examining a large number of states confirms that, although not absolute, this is the typical scenario: the low-energy spectrum is largely divided into odd integrable states and even chaotic states.
\section{Results for three particles \label{sec:results_3p}}
\medskip
We now turn to the case of $N=3$ identical bosons. In this case, the Bethe numbers are integers, and we truncate the basis to the range $-N_s \leq I_1 < I_2 < I_3 \leq N_s$ with $N_s = 17$. The dimension of the corresponding truncated Hilbert space is therefore $M=\binom{2N_s+1}{3}=6545$.

\begin{figure}
  \centering\includegraphics[width=0.7\linewidth]{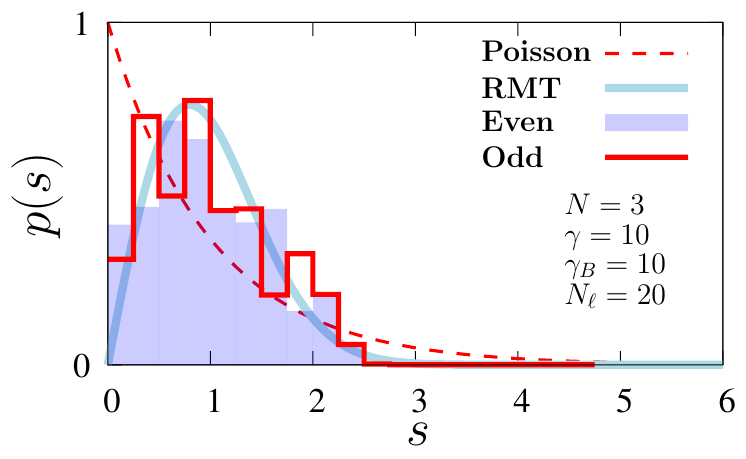}
  \caption{Level spacing distribution of the unfolded spectrum for three particles. The histogram is computed over the lowest $N_\ell=20$ energy levels for different $\gamma$ and $\gamma_B$ in the range [5-10]. In total 2200 levels are used to compute these histograms. The levels are sorted by parity and the numerical histograms are compared to the Poisson and Wigner--Dyson distributions. \label{fig_ps_3p}.}
\end{figure}

\begin{figure}
  \centering\includegraphics[width=0.7\linewidth]{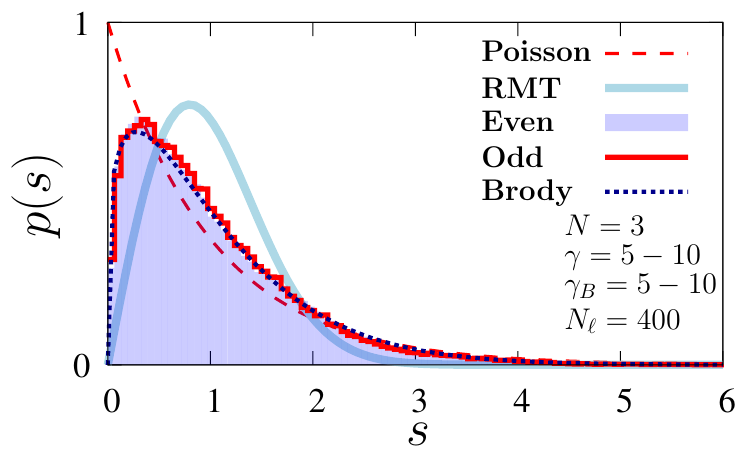}
  \caption{Level spacing distribution of the unfolded spectrum for three particles. The histogram is computed over the lowest $N_\ell=400$ energy levels for different $\gamma$ and $\gamma_B$ in the range [5-10]. In total 44000 levels are used to compute these histograms. The levels are sorted by parity and the numerical histograms are compared to the Poisson, Wigner--Dyson and Brody distributions (with $\beta=0.3715$) \cite{fitbrody}.\label{fig_ps_3p2} }
\end{figure}

The structure of the spectrum is similar to the two-particle case shown in Fig.~\ref{fig_spectrum_2p}. We focus on the regime where chaotic behavior is expected. The phenomenology is also analogous to the two-particle case, with the important difference that the odd-parity sector of the spectrum is no longer integrable. Figure~\ref{fig_ps_3p} shows the LSD in the low-energy sector of the spectrum and demonstrates that both odd- and even-parity states display clear signatures of chaos. However, when the high-energy part of the spectrum is included in the calculation of the LSD, strong deviations from random-matrix statistics appear. In this regime, the distribution is well described by a Brody distribution, as shown in Fig.~\ref{fig_ps_3p2}. Importantly, even and odd sectors share the exact same statistics which still present signatures of quantum chaos. 

When higher-energy levels are included, an important difference emerges between the $N=2$ and $N=3$ cases. Although both systems display a crossover from chaotic behavior at low energy to quasi-integrable statistics at high energy, this crossover occurs at significantly lower excitation numbers for $N=2$ than for $N=3$. The origin of this difference lies in the density of states, which increases with the number of particles. As a consequence, selecting the same number of eigenstates corresponds to much higher energies for $N=2$ than for $N=3$, where high-energy states are already largely insensitive to the barrier. This explains why Poisson-like statistics is recovered more rapidly in the two-particle case, while signatures of level repulsion persist over a broader range of levels for $N=3$.

\begin{figure}
 \centering \includegraphics[width=0.6\linewidth]{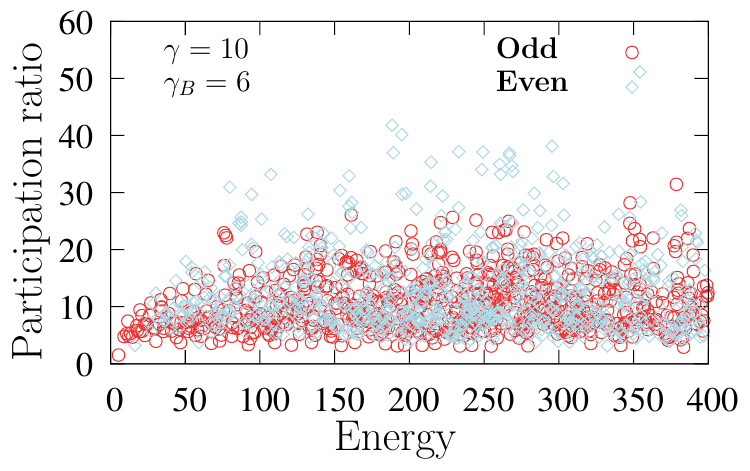}
  \caption{Participation ratio of the eigenstates of hamiltonian (\ref{eq_H_LL}) for $N=3$, $\gamma=10$ and $\gamma_B=6$. Red circles correspond to the odd parity states and blue diamonds to the even parity states.\label{fig_PR_3p} Energy is in units of $\hbar^2/mL^2$.}
\end{figure}

\begin{figure}
  \centering\includegraphics[width=0.6\linewidth]{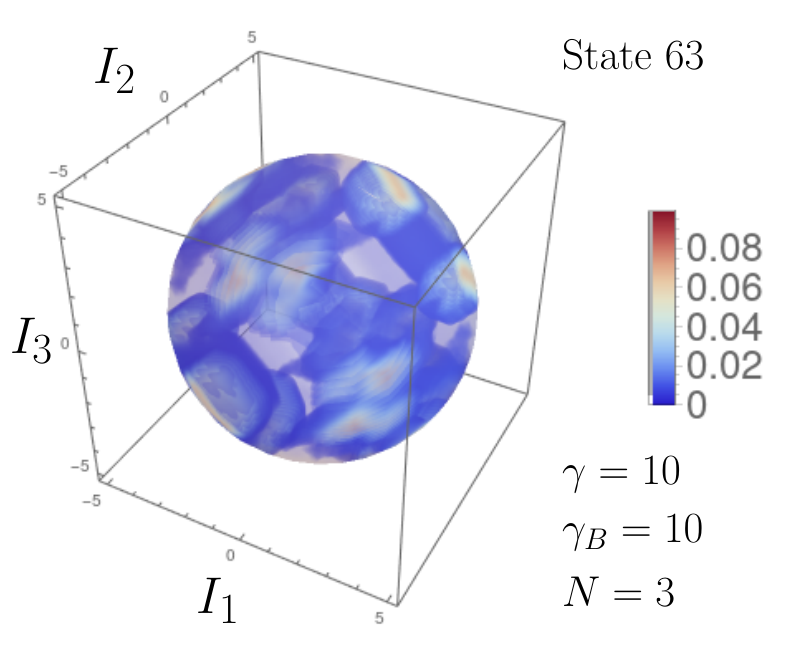}
  \caption{Modulus square of the wave function $|\alpha^{(n)}_{\vec \lambda}|^2$ of state $n=63$ in the Bethe space ($I_1,I_2,I_3$) with parameters $\gamma=10$ and $\gamma_B=10$. This eigenstate spreads uniformly over the constant energy shell defined as $E=\frac{\hbar^2}{2m}\left(\frac{2\pi}{L}\right)^2 (I_1^2+I_2^2+I_3^2)$ in Bethe space (here $E\simeq 50.51\hbar^2/mL^2$). \label{fig_state-crop.pdf} }
\end{figure}

As in the two-particle case, we also compute the participation ratio, shown in Fig.~\ref{fig_PR_3p}. In contrast, no clear distinction between states of different parity is observed. The participation ratio does not systematically grow with energy, as would be expected for fully chaotic states. Nevertheless, a substantial fraction of the eigenstates exhibits large participation ratios and spreads almost uniformly over the constant-energy shell, as illustrated in Fig.~\ref{fig_state-crop.pdf}. Taken together with the spectral statistics, these results demonstrate that the LL model coupled to a delta barrier undergoes a transition to quantum chaos.
\section{Discussion \label{sec:discussion}}
\medskip
Let's use the $N=2$ case as a paradigmatic example of the thermalization scenario we are going to suggest. For reference, let us consider the following classical stochastic model: our classical particles move freely between particle-particle and particle-barrier collisions; on a collision, particles undergo either a transmission or a reflection event, with the quantum-informed probabilities
\begin{align*}
&
\text{Prob}_{\text{transmission, particle-particle}} = |t|^2
\\
&
\text{Prob}_{\text{reflection, particle-particle}} = |r|^2
\\
&
\text{Prob}_{\text{transmission, barrier}} = |t_{B}|^2 
\\
&
\text{Prob}_{\text{reflection, barrier}} = |r_{B}|^2
\end{align*}
where
\begin{align}
\begin{split}
&
t = \frac{i |k_{\text{rel.}}|a}{1+(|k_{\text{rel.}}|a)^2}
\\
&
r = -\frac{1}{1+(|k_{\text{rel.}}|a)^2}
\\
&
t_{B} = \frac{i |k_{1,2}|a}{1+(|k_{1,2}|a_{B})^2}
\\
&
r_{B} = -\frac{1}{1+(|k_{1,2}|a_{B})^2}
\end{split}
\label{tr}
\end{align}
are the corresponding quantum amplitudes. Here, $  k_{1}=p_{1}/\hbar  $ ($  k_{2}=p_{2}/\hbar  $) is the wavenumber of particle 1 (2), $  p_{1}  $ and $  p_{2}  $ are the corresponding momenta, $  k_{\text{rel.}}=k_{2}-k_{1}  $ is the relative wavenumber. Likewise, the scattering lengths are $  a=\hbar/(m/2) g  $ for the particle-particle collisions and $  a_B=\hbar/m g_{B}  $ for the barrier. Observe that in this model, our particle-particle and particle-barrier interactions are only capable of either interchanging particle momenta, $  (p_{1},p_{2})\to (p_{2},p_{1})  $, or changing the sign of one of the two components, $  (p_{1},p_{2})\to (-p_{1},p_{2})  $ or $(p_{1},p_{2})\to (p_{1},-p_{2})$. If these transitions were the only ones allowed, the barrier would couple any given even-parity LL eigenstate to only one other state, leading to a participation ratio no greater than $2$. The same holds for the odd states. However, there exists a purely quantum-mechanical source of new values of particle momenta, namely diffraction. 

Scattering of two $\delta$-interacting distinguishable particles on a $  \delta  $-barrier, on a line, was analyzed in the classic book of Morse and Feshbach~\cite{book_morse_feshbach} (\S12.3). In the case of repulsive interactions, in addition to the expected classical channels listed above, they describe a process in which the particles collide in the vicinity of the barrier. The presence of the barrier temporarily lifts momentum conservation and allows particles to \emph{partially} exchange energy, still remaining on the equi-energy shell controlled by the initial energy. In the purely quantum domain, the book~\cite{book_morse_feshbach} reinterprets this effect as a circular wave emanating from the origin. This effect belongs to the realm of quantum-mechanical diffraction. The book goes further in identifying the ``sharp object'' that causes it. If one attempts to construct a solution as a sum of a finite number of plane waves, in each of the six domains separated by the interactions, one would find that the outgoing portions of these waves must have a discontinuity in their amplitude. The reason for this discontinuity is that the amplitude of the resulting wave depends on the order of collisions in the past. This discontinuity can be interpreted as a violation of the Yang-Baxter relation \cite{gaudin1983_book_english} (Sec.~10.3.2) for a three-body collision of two particles of finite mass and one massive particle. Interestingly, such a discontinuity does not appear in the outgoing wave that results solely from a sequence of transmission events.

The emergence of diffraction in the presence of a barrier can be illustrated using the \emph{eikonal approximation} \cite{Landay_EandM}. There, one attempts to use classical trajectories as the ``skeleton'' for a quantum eigenstate: the wavefunction is approximated by a phase factor
$  \exp[i S(\vec{r},\,\vec{r}_{0})/\hbar]  $ where
$S(\vec{r},\vec{r}_{0}) \equiv
\int_{\vec{r}_{0}}^{\vec{r}} \vec{p} \cdot d\vec{r}$ is the \emph{abbreviated action} between a point $  \vec{r}_{0}  $ on a chosen constant phase front of the incident wave and the coordinate of interest $  \vec{r}  $, with $  \vec{r}_{0}  $ being connected to $  \vec{r}  $ by a trajectory.

Interestingly, semitransparent zero-range objects can also be incorporated in the eikonal framework. Each allowed sequence of transmissions and reflections
must be treated separately, and the resulting quantum solutions should be added up. Each collision event leads to a multiplication of the resulting
wavefunction by the corresponding transmission or reflection amplitude \eqref{tr}.
The eikonal approximation can also be reinterpreted as an attempt to build the quantum solution as a superposition of plane waves with the classically allowed momenta; the approximate solution is allowed to have discontinuities that separate areas with different collision pre-histories. Such discontinuities signify the limits of the validity
of the eikonal approximation.

\begin{figure}
\centering \includegraphics[width=0.8\linewidth]{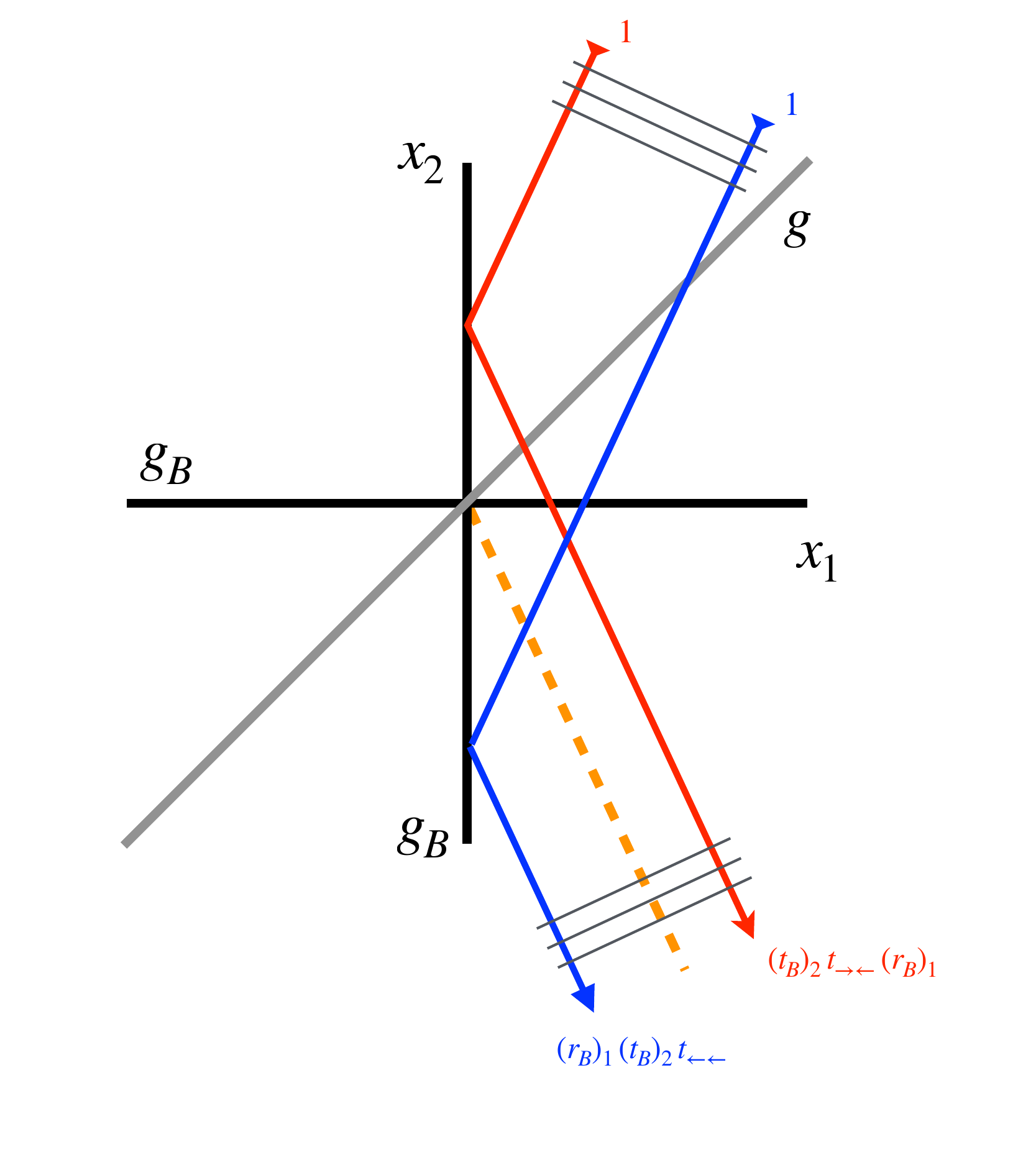}
\caption{An illustration of one of the sources of diffraction in a system consisting of two distinguishable particles interacting between themselves and with a barrier. The lines correspond to the particle-particle interaction line (grey), particle-barrier interaction line (black), trajectory 1 (red), trajectory 2 (blue), and the line along which a discontinuity in the wavefunction is predicted, at this order of approximation (orange, dashed). See explanations in text.}
\label{fig_diffraction}
\end{figure}

Fig.~\ref{fig_diffraction} provides an example of the appearance of diffraction in our system. We start from a two-body plane wave corresponding to both particle 1 and particle 2 approaching the barrier from the right, with particle 2 being faster than particle 1 and behind it. We will be interested in the outgoing two-particle wave where particle 1 moves to the right, and particle 2 moves to the left. Depending on the initial positions and velocities of the particles, two distinct scenarios are possible. According to scenario 1 (red trajectory), particle 1 bounces off the barrier, penetrates through particle 2 and escapes to positive infinity, while particle 2 continues towards the barrier, penetrates it and escapes to negative infinity. Scenario 2 (blue trajectory) is different. There, particle 1 is first penetrated by particle 2, who continues towards the barrier, penetrates it and escapes to negative infinity; finally, particle 1 gets reflected by the barrier and continues moving towards positive infinity.
Assume that trajectories 1 and 2 originate from the same incident wavefront of the quantum eigenstate that they model. Let them both propagate for the same amount of time, and, consequently, travel the same two-dimensional distance. The principle of least action will guarantee that the phase factor $  \exp[i S/\hbar]  $ will be the same for both trajectories. However, due to the differences between their collision sequences, the magnitudes of the resulting
waves---$  (t_{B})_{2} t_{\rightarrow\leftarrow}(r_{B})_{1}  $ and $  (r_{B})_1(t_{B})_{2}t_{\leftarrow\leftarrow}  $ respectively---will be different. Note that two points with different prehistories can be arbitrarily close, leading to a discontinuity (orange line at Fig.\ref{fig_diffraction}) in the \emph{predicted} value of the wavefunction.
In the full quantum-mechanical solution, the discontinuities are rectified by diffraction, and this is what we observe numerically.

Curiously, in instances where a Bethe Ansatz is available, the eikonal approximation becomes exact, leading to this solution. It will thus be instructive to analyze how the diffraction disappears for the spatially odd eigenstates of our system, which are accessible via the asymmetric Bethe Ansatz.
The odd solutions feature a node along the secondary diagonal, $  x_1 = -x_2  $. One can then place a hard wall along that diagonal: thanks to the node, this wall does not alter the solutions, and the problem can be solved separately on each side of the wall. Finally, as can be observed from Fig.~\ref{fig_diffraction}, such a wall forbids trajectories with different pre-histories from simultaneously coming close in space and propagating in the same direction. This prohibition prevents diffraction.
The above construction does not generalize to three or more particles. One can show that any node structure of the form $  x_i = -x_j  $ will either fail to shield the surfaces of discontinuity, require a node on the barrier itself, or contradict the bosonic nature of the particles.

Notice that the diffraction is a purely quantum effect: classically, the probability of two point-like particles colliding while in a zero volume region of space occupied by the barrier is zero. This explains the low participation ratio that we observed in Fig.~\ref{fig_PR_2p}. The participation ratio, i.e., the number of the unperturbed states participating in a given perturbed state translates, in the classical limit, to the range of the unperturbed canonical actions spanned by the perturbed trajectory divided by $(2\pi\hbar)^2$ (for two-dimensional systems). For the purely quantum effects, this range must vanish in the limit of $\hbar\to 0$. For this to happen, the participation ratio must grow slower than $(2\pi\hbar)^{-2}$; plateauing at a finite value is one of the options. For two particles, a typical value of the participation ratio appears to be about $4$ (Fig.~\ref{fig_PR_2p}). Tellingly, this is exactly the value identified~\cite{kurlberg2017_2947} in another purely diffractive, partially thermalizable system: \v{Seba} billiard ~\cite{seba1990}. In that model, a two-dimensional particle on a torus interacts with a point scatterer---an object with no classical analogue---making diffraction the sole source of thermalization. An upper bound on the participation ratio explains why the proposed diffraction-induced thermalization is essentially a low-energy effect. As energy increases, the number of two-body momentum states required to fill the microcanonical equi-energy shell grows. Eventually, this number exceeds the participation-ratio bound, leading to unpopulated ``holes'' on the microcanonical surface. Finally, in another respect, our system is more chaotic than the \v{Seba} billiard. Fig.~\ref{fig_ps_2p} demonstrates that for large level spacings, our system remains close to the random matrix theory prediction, exhibiting a Gaussian behavior. By contrast, it has been shown that \v{Seba} billiards retain an exponential tail~\cite{bogomolny1999_R1315,bogomolny2001,bogomolny2002} inherited from integrable systems.

Our system needs to be compared to another well-studied quantum chaotic system that exhibits stabilization of the participation ratio in some regimes: the quantum kicked rotor \cite{kicked_rotor}. There, for a sufficiently high kick frequency, the participation ratio stops growing, thus halting the exploration of the available phase space. However, while in our case the participation ratio is bounded from above by a constant number, in the kicked rotor it can be controlled by the kick strength.

\section{Conclusion\label{sec:conclusion}}
\medskip
We have investigated the emergence of quantum chaos in the Lieb--Liniger model of interacting bosons coupled to a localized delta barrier. While the model is integrable in the absence of impurities, the barrier induces diffraction and breaks integrability in a controlled manner. By analyzing the spectral statistics for two and three particles, we demonstrated that the low-energy spectrum exhibits signatures of Gaussian orthogonal ensemble statistics, in contrast to the Bohigas--Giannoni--Schmit conjecture which associates chaotic behavior with highly excited states. For two bosons, we found a striking separation between parity sectors: the odd sector remains integrable while the even sector is chaotic at low energies and tends towards quasi-integrable behavior at higher energies. For three bosons, both parity sectors show clear signs of chaos in the low-energy regime.

We argue that in generic one-dimensional systems consisting of particles of the same mass, diffraction is the fundamental mechanism for the onset of chaos. Beyond few-body physics, our findings open the way to exploring the role of diffraction in a quantum thermalization process along with the quantum dynamical phenomena such as transport, relaxation, and entanglement growth.

\section*{Acknowledgments}
We would like to acknowledge helpful discussions with A. Minguzzi. This work has benefited from the financial support of Agence Nationale de la Recherche under Grant No ANR-21-CE47-0009 Quantum-SOPHA, the Grant No ANR-23-PETQ-0001 Dyn1D France 2030, and the PRIN
2022 (2022R35ZBF) - PE2 - ``ManyQLowD'' as well as a one month invited professor grant for M. Olshanii from Universit\'e C\^ote d'Azur. M.O. was supported by the NSF Grant No. PHY-2309271.

\appendix
\section{Calculation of the matrix elements of the barrier in the Bethe basis\label{sec:appendix}}
\medskip
 In this Appendix, we will detail the derivation of matrix elements involving the contribution of the barrier potential, which the diagonalization of the Hamiltonian (\ref{eq_H_LL}) depends upon. These matrix elements are written in the Bethe basis $|\vec\lambda\rangle$ and read
\begin{align}
    C_{\{\lambda_k\},\{\mu_k\}}(\gamma_B)&=\gamma_B\langle\vec\lambda|\sum_{i=1}^N\delta(x_i)|\vec\mu\rangle,\\
    &=\gamma_B\int\prod_{j=1}^Ndx_j~\sum_{i=1}^N\delta(x_i)\Psi^*_{\{\lambda_k\}}(\{x_k\})\Psi_{\{\mu_k\}}(\{x_k\}).
\end{align}
\noindent We recall that the total Bethe wavefunction for identical bosons is defined as $\langle x_1,\dots,x_N|\vec\lambda\rangle=\Psi_{\{\lambda_j\}}(\{x_j\})$, which takes the following form
\begin{equation}
   \Psi_{\{\lambda_k\}}(\{x_k\})=\sum_{Q\in S_N} A_Q(\{\lambda_k\},c) \sum_{P\in S_N} \theta_P(\{x_k\})e^{i\sum_jx_{P(j)}\lambda_{Q(j)}}.
\end{equation}
\noindent where $S_N$ is the permutation group of dimension $N!$, $\theta_Q(\{x_k\})=\theta(x_{Q(1)}<\dots<x_{Q(N)})$ is the indicator function whose value is 1 in the sector Q defined as $x_{Q(1)}<\dots<x_{Q(N)}$ and 0 elsewhere. 
\noindent Putting this expression in the matrix element, we have
\begin{align}
    C_{\{\lambda_k\},\{\mu_k\}}(\gamma_B)&=\gamma_B\sum_{Q,Q'}A_{Q}A_{Q'}\int\prod_{j=1}^Ndx_j~\sum_{i=1}^N\delta(x_i)\nonumber\\
    &\times\sum_{P,P'}\theta_{P}(\{x_k\})\theta_{P'}(\{x_k\})e^{-i\sum_q(x_{P(q)}\lambda_{Q(q)}-x_{P'(q)}\mu_{Q'(q)})},
\end{align}
\noindent where $A_{Q}=A_{Q}{(\{\lambda_k\},c)}$, $A_{Q'}=A_{Q'}{(\{\mu_k\},c)}$. Due to the product of $\theta$ functions, the previous formula is only different from 0 when $P=P'$, which leads to
\begin{align}
    C_{\{\lambda_k\},\{\mu_k\}}(\gamma_B)=\gamma_B
\sum_{Q,Q'}A_{Q}A_{Q'}\sum_{P}\int_{\Gamma_P}\prod_{j=1}^Ndx_j~\sum_{i=1}^N\delta(x_i)e^{-i\sum_qx_{P(q)}(\lambda_{Q(q)}-\mu_{Q'(q)})},
\end{align}
\noindent with $\Gamma_P$ delimiting the integration domain $x_{P(1)}<\dots<x_{P(N)}$. Let us rewrite the different sectors by discriminating the Dirac delta's variable
\begin{align}
    C_{\{\lambda_k\},\{\mu_k\}}(\gamma_B)&=\gamma_B\sum_{Q,Q'}A_{Q}A_{Q'}\sum_{P'}\sum_{k=1}^N\int_{x_k<\Gamma_{P'}}\prod_{j}dx_j~\sum_{i=1}^N\delta(x_i)\nonumber\\
    &\times e^{-ix_{k}(\lambda_{Q(1)}-\mu_{Q'(1)})}e^{-i\sum_{q=2}^Nx_{P'(q)}(\lambda_{Q(q)}-\mu_{Q'(q)})},
\end{align}
\noindent where $P'$ represent the permutations of the elements $\{2,\dots,N\}$ and $x_k<\Gamma_{P'}$ means $x_k<x_{P'(2)}<\dots<x_{P'(N)}$.
To understand how these integrals behave, let's dissect the $N=3$ case and consider the integrals of the form
\begin{equation}
    \int_{x_i<x_j<x_k}dx_idx_jdx_k~\delta(x_n)e^{-ix_il_i}e^{-ix_jl_j}e^{-ix_kl_k},
\end{equation}
\noindent with $i,j,k,n=1,2,3$ and $l_m=\lambda_{Q(m)}-\mu_{Q'(m)}$. This integral can be explicitly computed while reshaped into the following form
\begin{equation}
    \int_0^Ldx_i~e^{-ix_il_i}\int_{x_i}^Ldx_j~e^{-ix_jl_j}\int_{x_j}^Ldx_k~\delta(x_n)e^{-ix_kl_k}.
\end{equation}
\noindent We now have three cases to treat depending on $n$'s value. If $n=i$, the integral in $x_i$ becomes irrelevant, and we are left with
\begin{equation}
    \int_{0}^Ldx_j~e^{-ix_jl_j}\int_{x_j}^Ldx_k~\delta(x_n)e^{-ix_kl_k},
\end{equation}
\noindent the two other cases are more subtle. If $n=j$ the sector can be separated as $0<x_i<\epsilon$ and $\epsilon<x_k<L$ for $\epsilon\rightarrow0$. The first part has no spatial extent, effectively canceling the integral. The same reasoning can be applied to the remaining case, leading to the ensuing result
\begin{equation}
    \int_{x_i<x_j<x_k}dx_idx_jdx_k~\delta(x_n)e^{-ix_il_i}e^{-ix_jl_j}e^{-ix_kl_k}=\int_{0}^Ldx_j~e^{-ix_jl_j}\int_{x_j}^Ldx_k~e^{-ix_kl_k}.
\end{equation}
\noindent By going back to the $N$-body problem, the matrix elements can now be written as
\begin{align}
    C_{\{\lambda_k\},\{\mu_k\}}(\gamma_B)=\gamma_B
\sum_{Q,Q'}A_{Q}A_{Q'}\sum_{P'}\int_{\Gamma_{P'}}\prod_jdx_j~\sum_{i=1}^N\delta(x_i)e^{-i\sum_qx_{P'(q)}(\lambda_{Q(q)}-\mu_{Q'(q)})},
\end{align}
\noindent which can be further simplified since all sectors are equivalent. As such, the final form of these elements reads
\begin{equation}
    C_{\{\lambda_k\},\{\mu_k\}}(\gamma_B)=\gamma_B\sum_{Q,Q'}A_{Q,Q'}S_{Q,Q'},
\end{equation}
\noindent with the reduced matrix elements
\begin{align}
    S_{Q,Q'}&=N!\int_0^Ldx_2e^{-ix_{2}(\lambda_{Q(2)}-\mu_{Q'(2)})}\prod_{j=3}^N\int_{x_{j-1}}^Le^{-ix_{j}(\lambda_{Q(j)}-\mu_{Q'(j)})}.
    \label{eq:A_alpha R}
\end{align}
In the $N=2$ and $N=3$ cases detailed in the main, these reduced matrix elements have been exactly computed and can be put in the following form
\begin{align}
    S_{Q,Q'}&=2i\dfrac{e^{-iL(\lambda_{Q(2)}-\mu_{Q'(2)})}}{\lambda_{Q(2)}-\mu_{Q'(2)}},
\end{align}
for the case $N=2$, and
\begin{align}
    S_{Q,Q'}&=6L\Bigl[\dfrac{e^{-iL(\lambda_{Q(3)}-\mu_{Q'(3)})}}{(\lambda_{Q(2)}-\mu_{Q'(2)})(\lambda_{Q(3)}-\mu_{Q'(3)})}\nonumber\\
    &-\dfrac{e^{-iL(\lambda_{Q(2)}-\mu_{Q'(2)}+\lambda_{Q(3)}-\mu_{Q'(3)})}}{(\lambda_{Q(2)}-\mu_{Q'(2)})(\lambda_{Q(2)}-\mu_{Q'(2)}+\lambda_{Q(3)}-\mu_{Q'(3)})}\nonumber\\
    &-\dfrac{1}{(\lambda_{Q(3)}-\mu_{Q'(3)})(\lambda_{Q(2)}-\mu_{Q'(2)}+\lambda_{Q(3)}-\mu_{Q'(3)})}\Bigr]
\end{align}
for the case $N=3$.
%
\bibliographystyle{apsrev}

\begin{thebibliography}{199}
\bibitem{BGS1984} O. Bohigas, M. J. Giannoni, and C. Schmit, \href{https://doi.org/10.1103/PhysRevLett.52.1}{Phys. Rev. Lett. {\bf 52}, 1 (1984)}.
\bibitem{RMTreview} T. Guhr, A. M\"uller-Groeling, and H. A. Weidenm\"uller, \href{https://doi.org/10.1016/S0370-1573(97)00088-4}{Phys. Rep. {\bf 299}, 189 (1998)}.
\bibitem{RMTreview2} O. Bohigas and H. A. Weidenmueller, \href{https://doi.org/10.1146/annurev.ns.38.120188.002225}{Ann. Rev. Nucl. Part. Sci. {\bf 38}, 421 (1988)}.
\bibitem{LiebLiniger}
E. H. Lieb, and W. Liniger, \href{https://doi.org/10.1103/PhysRev.130.1605}{Phys. Rev. {\bf 130}, 1605 (1963)}.
\bibitem{Paredesetal}
 B. Paredes, A. Widera, V. Murg, O. Mandel, S. F\"olling, I. Cirac, G. V. Shlyapnikov, T. W. H\"ansch, and I. Bloch, \href{https://doi.org/10.1038/nature02530}{Nature {\bf 429}, 227 (2004)}.
\bibitem{Kinoshita2004} T. Kinoshita, T. Wenger, and D. S. Weiss, \href{https://www.science.org/doi/10.1126/science.1100700}{Science {\bf 305}, 1125 (2004)}.
\bibitem{Kinoshita2005} T. Kinoshita, T. Wenger, and D. S. Weiss, \href{https://doi.org/10.1103/PhysRevLett.95.190406}{Phys. Rev. Lett. {\bf 95}, 190406 (2005)}.
\bibitem{CazalillaCitroGiamarchiOrigancRigol} M. A. Cazalilla, R. Citro, T. Giamarchi, E. Orignac and M. Rigol, \href{https://doi.org/10.1103/RevModPhys.83.1405}{Rev. Mod. Phys. {\bf 83}, 1405 (2011)}.
\bibitem{WangProsen2025} J. Wang, T. Prosen, and G. Casati, \href{https://link.aps.org/doi/10.1103/lkjy-xd53}{Phys. Rev. E. {\bf 112}, L032201 (2025)}.
\bibitem{Girardeau} M. Girardeau, \href{https://doi.org/10.1063/1.1703687}{J. Math. Phys. {\bf 1}, 516 (1960)}.
\bibitem{Korepinetal} V. E. Korepin, N. M. Bogoliubov, and A. G. Izergin, {\it Quantum Inverse Scattering Method and Correlation Functions} (Cambridge University Press, Cambridge, 1983).
\bibitem{Slavnov} N. A. Slavnov, \href{https://doi.org/10.1007/BF01016531}{Theoret. and Math. Phys. {\bf 79}, 502 (1989)}.
\bibitem{deNardisPanfil} J. De Nardis and M. Panfil \href{https://dx.doi.org/10.1088/1742-5468/2015/02/P02019}{J. Stat. Mech. P02019 (2015)}.
\bibitem{brody1973} T. A. Brody, {\it A statistical measure for the repulsion of energy levels}, \href{https://doi.org/10.1007/BF02727859}{Lett. Nuovo Cimento {\bf 7}, 482 (1973)}.
\bibitem{Izrailev2021} S. Mailoud, F. Borgonovi, F.~M. Izrailev, \href{https://journals.aps.org/pre/abstract/10.1103/PhysRevE.104.034212}{Phys. Rev. E {\bf 104}, 034212 (2021)}.
\bibitem{Jackson2025} M. Olshanii, M. Albert, G. Aupetit-Diallo, P. Vignolo, and S.~G. Jackson, \href{https://doi.org/10.21468/SciPostPhysCore.8.4.083}{SciPost Phys. Core {\bf 8}, 083 (2025)}.
\bibitem{fitbrody} We fitted our data to the Brody distribution using \texttt{gnuplot} and a least-squares minimization procedure. The histograms were constructed using 21 bins for $N=2$ over the range $s \in [0,5]$ and 123 bins for $N=3$ over the range $s \in [0,8]$.
\bibitem{book_morse_feshbach} P.M. Morse and H. Feshbach, {\it Methods of theoretical physics} (McGraw-Hill, Tokyo, 1953).
\bibitem{Landay_EandM} L. D. Landau and E.M. Lifshitz,
{\it The Classical Theory of Fields: v.~2 (Course of Theoretical Physics)} (Butterworth-Heinemann, Oxford, 1980)
\bibitem{gaudin1983_book_english}
Michel Gaudin and Jean-S\'{e}bastien Caux, {\it The Bethe wavefunction},
(Cambridge University Press, Cambridge, 2014).
\bibitem{kurlberg2017_2947}
P. Kurlberg and H. Uebersch\"{a}r, {\it Superscars in the Šeba billiard}, \href{https://doi.org/10.4171/jems/732}{J. Eur. Math. Soc.
{\bf 19}, 2947 (2017)}.
\bibitem{seba1990} P. \v{S}eba, {\it Wave chaos in singular quantum billiard}, \href{https://doi.org/10.1103/PhysRevLett.64.1855}{Phys. Rev. Lett.
{\bf 64}, 1855 (1990)}.
\bibitem{bogomolny1999_R1315} E. B. Bogomolny, U. Gerland, and C. Schmit, {\it Models of intermediate spectral statistics},
\href{https://doi.org/10.1103/PhysRevE.59.R1315}{Phys. Rev. E {\bf 59}, R1315 (1999)}.
\bibitem{bogomolny2001} E. Bogomolny, U. Gerland, and C. Schmit, {\it Singular statistics}, \href{https://doi.org/10.1103/PhysRevE.63.036206}{Phys. Rev. E {\bf 63}, 036206 (2001)}.
\bibitem{bogomolny2002} E. Bogomolny, O. Giraud, and C. Schmit, {\it Nearest-neighbor distribution for singular
billiards}, \href{https://doi.org/10.1103/PhysRevE.65.056214}{Phys. Rev. E {\bf 65}, 056214 (2002)}.
\bibitem{kicked_rotor} M. S. Santhanam, Sanku Paul, and J. Bharathi Kannan, {\it Quantum kicked rotor and its variants: Chaos, localization and beyond}, \href{https://doi.org/10.1016/j.physrep.2022.01.002}{Physics Reports {\bf 956}, 1–122 (2022)}.

\end{thebibliography}

\end{document}